\documentclass[]{llncs}

\usepackage{amsfonts}
\usepackage{color}
\usepackage{amsmath}

\usepackage{xargs}
\usepackage{ifthen}

\usepackage{longtable}
\usepackage{booktabs}
\usepackage{multirow}
\usepackage{rotating}
\usepackage{microtype}

\usepackage{paralist}

\usepackage{multirow}
\usepackage[bookmarks=false]{hyperref}

\newcommand{\myparagraph}[1]{\par\noindent\vspace{0.5mm}{\bf#1}}

\title{HordeQBF: A Modular and Massively Parallel QBF Solver\thanks{This
article will appear in the \textbf{proceedings} of the \emph{19th
International Conference on Theory and Applications of Satisfiability Testing
(SAT)}, LNCS, Springer, 2016.}}
\author{
	Tom\'{a}\v{s} Balyo\inst{1}\thanks{Supported by DFG project SA 933/11-1.} 
        \and 
        Florian Lonsing\inst{2}\thanks{Supported by the Austrian Science Fund (FWF) under grant S11409-N23.}
      }
\institute{
Karlsruhe Institute of Technology (KIT) \\ Karlsruhe, Germany 
\and 
Knowledge-Based Systems Group, Vienna University of Technology \\ Vienna, Austria
}

\begin{document}
\pagestyle{plain}

\maketitle

\begin{abstract}
The recently developed massively parallel satisfiability (SAT) solver HordeSAT
was designed in a modular way to allow the integration of any sequential
CDCL-based SAT solver in its core.  We integrated the QCDCL-based quantified
Boolean formula (QBF) solver DepQBF in HordeSAT to obtain a massively parallel
QBF solver---HordeQBF. In this paper we describe the details of this integration and report on
results of the experimental evaluation of HordeQBF's performance. 
HordeQBF achieves superlinear average and median speedup on the 
hard application instances of the 2014 QBF Gallery.
\end{abstract}


\section{Introduction}
\label{sec:introduction}

HordeSAT~\cite{DBLP:conf/sat/BalyoSS15} is a modular massively parallel SAT
solver which allows the integration of any sequential CDCL-based SAT solver in
its core.  This enables the transfer of advancements in CDCL SAT solving to a
parallel setting. Experiments showed that HordeSAT can achieve superlinear
average speedup on \nolinebreak hard \nolinebreak benchmarks.

The logic of quantified Boolean formulas (QBFs) extends SAT by explicit
quantification of propositional variables. Problems in complexity classes
beyond NP, particularly PSPACE-complete problems in domains like, e.g., formal
verification, reactive synthesis, or planning, can naturally be encoded as
QBFs.

QBF solvers based on QCDCL, the QBF-specific
variant of CDCL, apply techniques similar to
CDCL SAT solvers.  Thanks to this fact, it is possible to replace the SAT
solver in the core of HordeSAT by \emph{any} QCDCL QBF solver. Thereby,
it is not necessary to change the framework of HordeSAT which controls
the sharing of learned information and the execution of the core solver instances.

We integrated the latest public version 5.0 of the QCDCL-based solver
DepQBF~\cite{DBLP:conf/lpar/LonsingBBES15} in HordeSAT to obtain the massively
parallel QBF solver HordeQBF.  We present the implementation of HordeQBF,
which is not tailored towards the use of DepQBF as a core solver, and evaluate
its scalability on a computer cluster with 1024 processor cores. Experiments using
the application benchmarks of the 2014 QBF Gallery show that HordeQBF achieves
superlinear average and median speedup for hard instances.


\section{Preliminaries}
\label{sec:preliminaries}

We consider closed QBFs $\psi := \Pi.\phi$ in \emph{prenex CNF (PCNF)}
consisting of a quantifier-free CNF $\phi$ over a set $V$ of variables and a \emph{quantifier
  prefix} $\Pi := Q_1v_1 \ldots Q_nv_n$ in which $Q_i\in\{\exists, \forall\}$ and $v_i\in V$. 
QBF solving with clause and cube learning
(\emph{QCDCL})~\cite{DBLP:journals/jair/GiunchigliaNT06,DBLP:conf/tableaux/Letz02,DBLP:conf/cp/ZhangM02},
also called \emph{constraint learning}, 
is a generalization of \emph{conflict-driven clause learning (CDCL)} for
SAT. The variables in a PCNF $\psi$ are assigned by
\emph{decision making}, \emph{unit propagation}, and \emph{pure literal
detection}. Assignments by decision making have to follow the prefix ordering
from left to right. If a clause is falsified under the current assignment $A$,
then a \emph{learned clause} $C$ is derived from $\psi$ by
\emph{Q-resolution}~\cite{DBLP:journals/iandc/BuningKF95} and added
conjunctively to $\psi$. If all clauses are satisfied under $A$, then a
\emph{learned cube} is constructed from $A$ and added disjunctively to
$\psi$. Learned cubes may also be derived by \emph{term
resolution}~\cite{DBLP:journals/jair/GiunchigliaNT06}, a variant of
Q-resolution applied to previously learned cubes. After a new clause or
cube has been learned, assignments are retracted during \emph{backtracking}. 
QCDCL terminates if and only if the \emph{empty clause
(resp. cube)} is derived during learning, indicating that $\psi$ is unsatisfiable
(resp. satisfiable).

\section{Related Work}

Approaches to parallel QBF solving are based on
shared and distributed memory architectures.  PQSolve~\cite{FMS00} is
an early parallel DPLL~\cite{DBLP:conf/aaai/CadoliGS98} solver without knowledge sharing. It
comes with a dynamic master/slave framework implemented using the
message passing interface (MPI)~\cite{gropp1996mpi}. Search space is partitioned among
master and slaves by variable assignments.
QMiraXT~\cite{DBLP:conf/mbmv/LewisSB09} is a multithreaded QCDCL
solver with search space partitioning.  PAQuBE~\cite{LSBMNG11} is an
MPI-based parallel variant of the QCDCL solver
QuBE~\cite{DBLP:journals/jsat/GiunchigliaMN10}. Clause and cube
sharing in PAQuBE can be adapted dynamically at run time. 
Search space is partitioned like in the SAT solver PSATO
based on \emph{guiding paths}~\cite{DBLP:journals/jsc/ZhangBH96}.  The
MPI-based solver MPI\-DepQBF~\cite{DBLP:conf/sat/JordanKLS14} implements
a master/worker architecture without knowledge sharing. A worker
consists of an instance of the QCDCL solver
DepQBF~\cite{DBLP:journals/jsat/LonsingB10}. The master balances the
workload by generating subproblems defined by variable assignments
(assumptions), which are solved by the workers.  Parallel solving
approaches have also been presented for quantified
CSPs~\cite{DBLP:conf/ictai/VautardLH10} and
non-PCNF~QBFs~\cite{DBLP:conf/ieeehpcs/MotaNS10}.

HordeQBF is a parallel portfolio solver with clause and cube
sharing. Whereas sequential portfolio solvers like
AQME~\cite{DBLP:journals/jsat/PulinaT10} include different QBF
solvers, HordeQBF integrates instances of the same QCDCL solver
(i.e., DepQBF). Unlike MPIDepQBF, HordeQBF does not rely on search
space partitioning. Instead, the parallel instances of DepQBF are
diversified by different parameter settings.


\section{The HordeSAT Parallelization Framework}

HordeSAT is a portfolio SAT solver with clause sharing~\cite{DBLP:conf/sat/BalyoSS15}. 
It can be viewed as a multithreaded program running several instances of a sequential SAT solver
and communicating via MPI 
with other instances of the same program. 

The parallelization framework has three main tasks: to ensure that the core solvers are diversified,
to handle the clause exchange, and to stop all the solvers when one of them has solved the problem. To
communicate with the core solvers it uses an API which is described in detail in the
HordeSAT paper~\cite{DBLP:conf/sat/BalyoSS15}. Since the HordeQBF interface is 
identical, we only briefly list the most relevant methods:
\myparagraph{\tt\bf void diversify(int rank, int size):}
This method tells the core solver to diversify its settings. The specifics of diversification
are left to the solver. The description for DepQBF is given in the following section.
\myparagraph{\tt\bf void addLearnedClause(vector$<$int$>$ clause):}
This method is used to import learned clauses (and cubes) received from other solvers of the portfolio.
\myparagraph{\tt\bf void setLearnedClauseCallback(LCCallback* callback):}
This method sets a callback class that will process the clauses (and cubes) shared by this solver.


\section{QBF Solver Integration}

In parallel QCDCL-based QBF solving, learned cubes may be shared among the
solver instances in addition to learned clauses. Although HordeSAT does not
provide API functions dedicated to cube sharing, its available API readily
supports it. We describe the integration of the QCDCL-based QBF
solver DepQBF\footnote{\url{http://lonsing.github.io/depqbf/}} in HordeQBF,
which applies to any QCDCL-based QBF solver.

We rely on version 5.0 of DepQBF which comes with a dynamic variant of
\emph{blocked clause elimination (QBCE)}~\cite{DBLP:conf/lpar/LonsingBBES15}
for advanced cube learning.  QBCE allows to eliminate redundant clauses from a
PCNF~\cite{DBLP:journals/jair/HeuleJLSB15}. Dynamic QBCE is applied frequently
during the solving process. If all clauses in the PCNF are satisfied under the
current assignment or removed by QBCE, then a cube is learned.

DepQBF features a sophisticated analysis of variable dependencies in a
PCNF~\cite{DBLP:journals/jsat/LonsingB10,DBLP:conf/sat/LonsingB10} to relax
the linear ordering of variables in the prefix. 
For the experiments in this paper, however,
we disabled dependency analysis for both HordeQBF and the sequential
variant of DepQBF since the use of
dependency information causes run time overhead (during clause/cube learning)
in addition to overhead already caused by dynamic blocked clause elimination
(QBCE) \cite{DBLP:conf/lpar/LonsingBBES15}.

We modified DepQBF as follows to integrate it in HordeQBF. Learned constraints
are exported to the master process right after they have been
learned. The master does not distinguish between learned clauses and cubes but
treats them as sorted lists of literals. We add special marker literals to
learned clauses and cubes to distinguish between them at the time when the
master provides the workers with sets of shared learned constraints.

In DepQBF we check whether shared constraints are available for import after a
restart has been carried out. To this end, we modified the restart policy of
DepQBF to always backtrack to decision level zero. This is different from the
original restart policy of DepQBF~\cite{DBLP:journals/jsat/LonsingB10}, where
the solver backtracks to higher decision levels depending on the current
assignment. After a restart, available shared constraints are imported, the
watched data structures are updated, and QCDCL continues by propagating unit
literals resulting from imported constraints.

\begin{figure}[!t]
\centering
\includegraphics[width=1\linewidth]{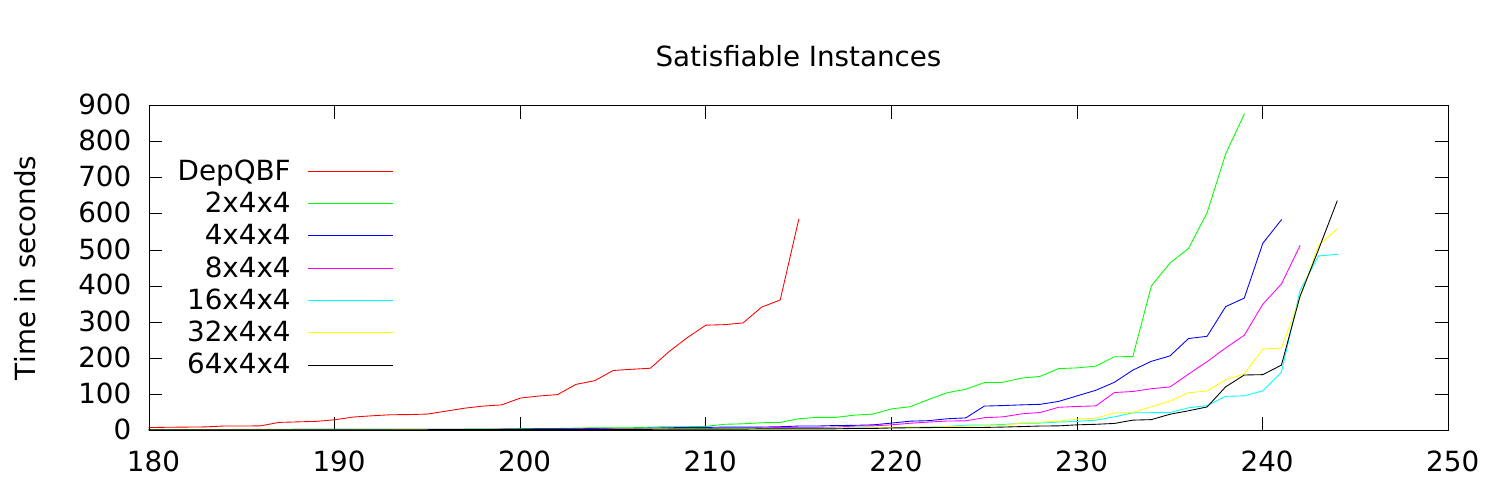}
\includegraphics[width=1\linewidth]{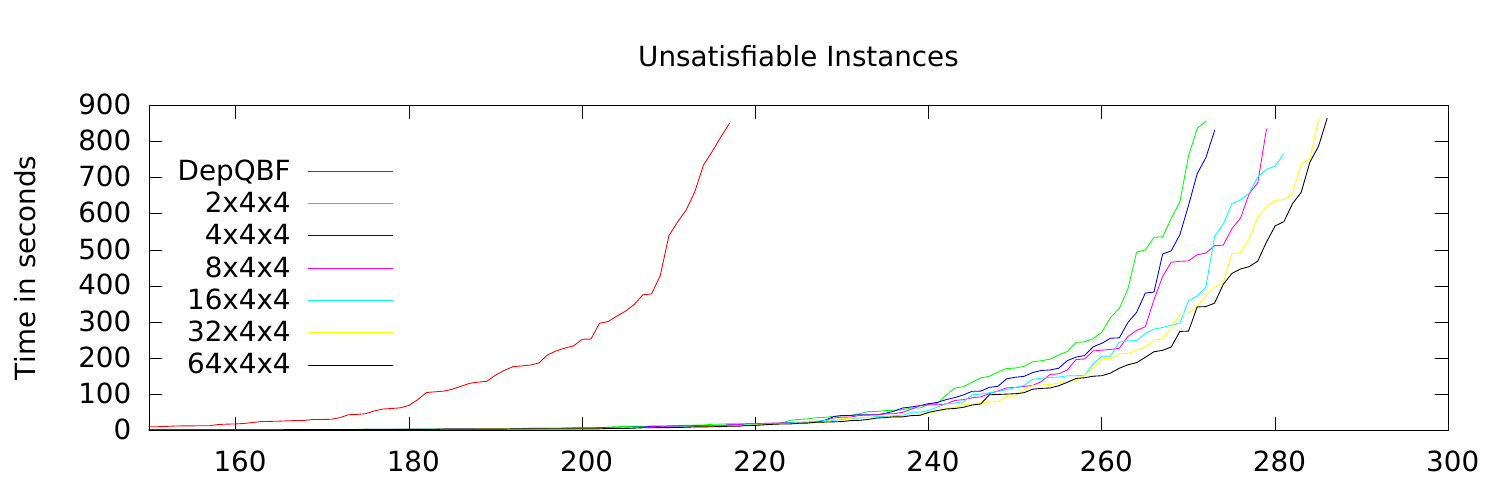}
\includegraphics[width=1\linewidth]{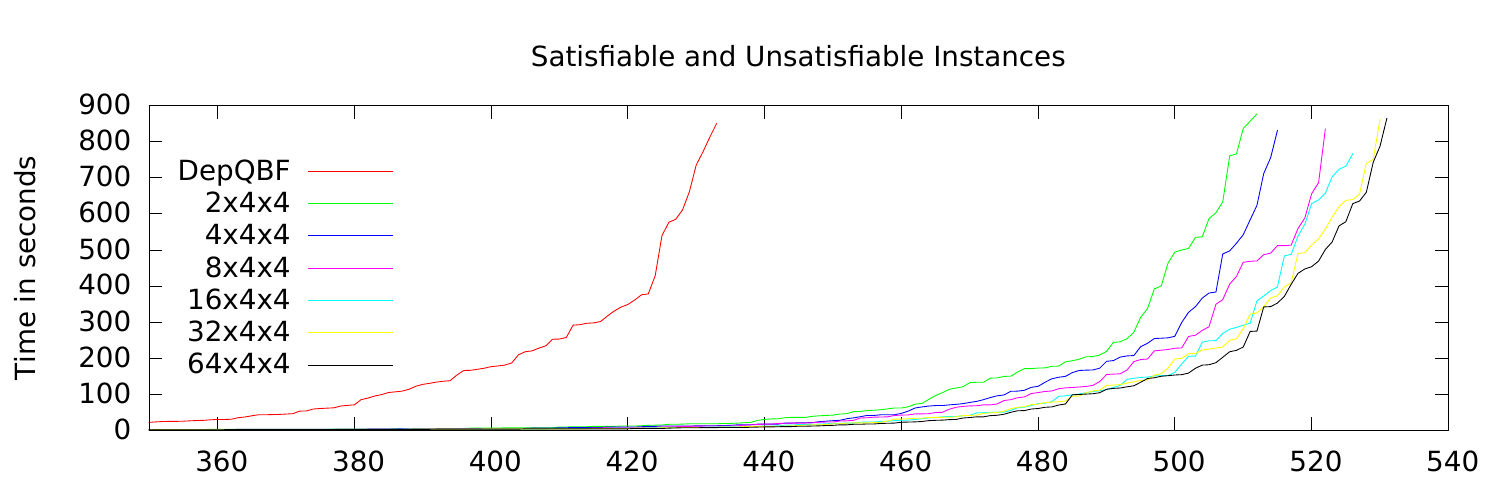}
\caption{Cactus plots for the benchmarks solved under 900 seconds by DepQBF and various configurations
of HordeQBF. The left regions of the plots (containing easy instances) are omitted.}
\label{fig_cact}
\end{figure}

Every instance of DepQBF receives a random seed from the master and
diversifies the solving process as follows.  The values of variables in the
\emph{assignment cache}~\cite{DBLP:conf/sat/PipatsrisawatD07} are initialized
at random. In general, decision variables are assigned to the cached value (if
any). The assignment cache is updated with values assigned by unit propagation
and pure literal detection. As an effect of random initialization, the first
value assigned to a decision variable is always a random value.  Parameters of
\emph{variable activity scaling} are set at random. DepQBF implements variable
activities similar to MiniSAT~\cite{DBLP:conf/sat/EenS03}. Additionally, the
amount (percentage) of learned constraints that are removed periodically is
initialized at random. DepQBF stores learned clauses and cubes in separate
lists with certain capacities. If a list has been filled during learning then
less frequently used constraints are removed and the capacity of the list is
increased.  DepQBF implements a \emph{nested restart scheme} similar to
PicoSAT~\cite{DBLP:journals/jsat/Biere08}, the parameters of which are
randomly selected.  Variants of \emph{dynamic
QBCE}~\cite{DBLP:conf/lpar/LonsingBBES15} are enabled at random, including
switching off dynamic QBCE at all, or applying QBCE only as a preprocessing or
inprocessing step. Finally, applications of \emph{long-distance
resolution}~\cite{DBLP:journals/fmsd/BalabanovJ12,DBLP:conf/iccad/ZhangM02},
an extension of traditional Q-resolution~\cite{DBLP:journals/iandc/BuningKF95}
used to derive learned constraints, are toggled at random.


\section{Experimental Evaluation}
\label{sec-exp}
To examine our portfolio-based parallel QBF solver HordeQBF we performed experiments using
all the 735 benchmark problems from the application track of the 2014 QBF Gallery
\cite{qbfgal14}. We compared HordeQBF with DepQBF, which is the QBF solver in the core of HordeQBF.

\begin{table}[t]
\renewcommand{\arraystretch}{1.15}
\renewcommand\tabcolsep{4.5pt}
\begin{center}

\begin{tabular}{r || r | r | r  r  r |  r  r  r  r}
Core~~ & Parallel & Both~ & \multicolumn{3}{c|}{Speedup All} & \multicolumn{4}{c}{Speedup Big} \\
Solvers & Solved~ & Solved & Avg. & Tot. & Med. & Avg. & Tot. & Med. & Eff.\\
\hline
2$\times$4$\times$4 & 513 & 483 & 622 & 107.30 & 0.82 & 3328 & 127.36 & 303.26 & 9.48\\
4$\times$4$\times$4 & 516 & 484 & 667 & 137.36 & 0.92 & 3893 & 176.27 & 458.34 & 7.16\\
8$\times$4$\times$4 & 523 & 492 & 748 & 128.35 & 0.96 & 4655 & 175.26 & 553.53 & 4.32\\
16$\times$4$\times$4 & 527 & 493 & 754 & 140.37 & 0.96 & 5154 & 236.18 & 1449.28 & 5.66\\
32$\times$4$\times$4 & 531 & 496 & 780 & 132.41 & 0.96 & 6282 & 269.87 & 2461.84 & 4.81\\
64$\times$4$\times$4 & 532 & 496 & 762 & 141.99 & 0.89 & 6702 & 307.29 & 2557.54 & 2.49\\
\end{tabular}

\end{center}
\caption{The speedup of HordeQBF configurations relative to DepQBF.
The second column is the number of instances solved by HordeQBF, the third is
the number of instances solved by both DepQBF (in 50000s) and the HordeQBF (in 900s).
The following six columns contain the average, total, and median speedups for either all the instances
solved by HordeQBF or only big instances (solved after 10$\times$\#cores seconds by DepQBF). The last
column is the parallel efficiency (median speedup/\#cores).}
\label{tab-speedup}
\end{table}

\begin{figure}[!t]
\centering
\includegraphics[width=1\linewidth]{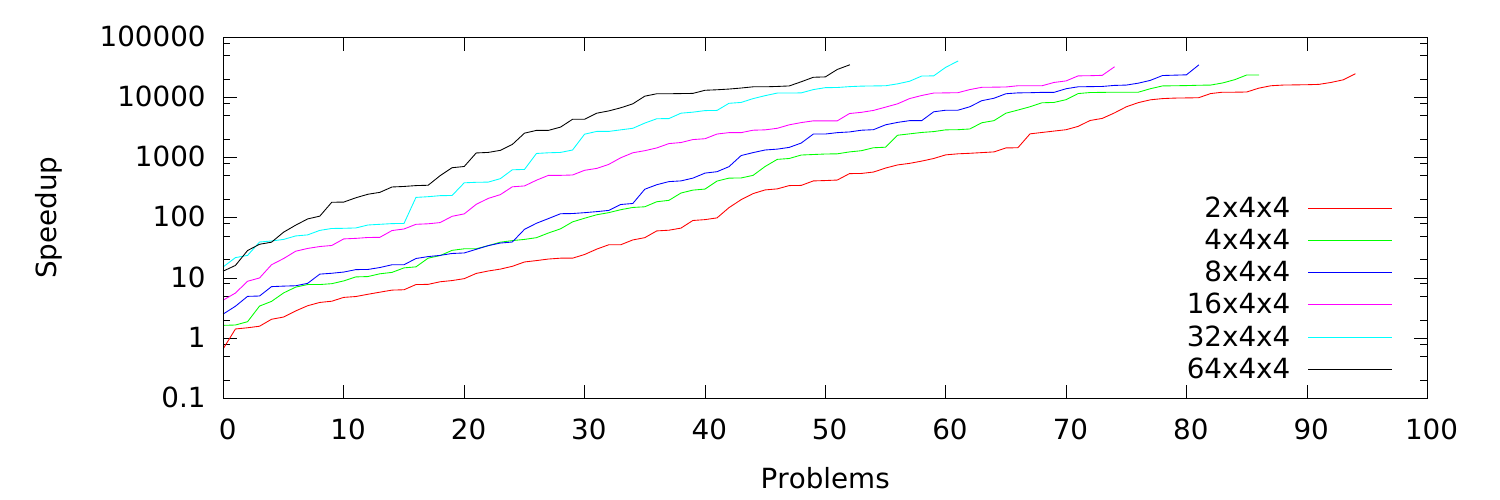}
\caption{Distribution of speedups on the ``big instances'' (solved after 10$\times$\#cores 
seconds by DepQBF -- the data corresponding to Columns~7--9 of Table~\ref{tab-speedup}).}
\label{fig_speedup}
\end{figure}

The experiments were run on a cluster with nodes having two octa-core 2.6 GHz Intel Xeon E5-2670 
processors (Sandy Bridge) and 64 GB of main memory.
Each node has 16 cores and we used 64 nodes which amounts in the total of 1024 cores.
The nodes communicate using an InfiniBand 4X QDR Interconnect and use
the SUSE Linux Enterprise Server 11 (x86\_64) (patch level~3) operating system.
HordeQBF was compiled using the icpc compiler version 15.0.2. The complete source code
and detailed experimental results are available at \url{http://baldur.iti.kit.edu/hordesat/}.

We ran experiments using 2, 4, \ldots, 64 cluster nodes. On each node we ran four processes 
with four threads each, which amounts to 16 core solver (DepQBF) instances per node.
The results are summarized in Figure~\ref{fig_cact} using cactus plots.
We can observe that increasing the number of cores is beneficial for both SAT and UNSAT instances
since the number of solved instances steadily increases and runtimes are reduced.

However, it is not easy to see from a cactus plot whether the additional performance is a
reasonable return on the invested hardware resources. Therefore we include Table~\ref{tab-speedup}
in order to quantify the overall scalability of HordeQBF.
We compute speedups for all the instances solved by the parallel solver.
We ran DepQBF with a time limit $T=50\,000s$ and for the instances it did not solve
we use the runtime of $T$ in speedup calculation. The parallel
configurations have a time limit of $900s$.
Columns 4, 5, and 6 of Table~\ref{tab-speedup} show the average, total (sum of sequential 
runtimes divided by the sum of parallel runtimes) and median speedup values respectively.
While the average and total speedup values are high, the median speedup is below one.

Nevertheless, these figures treat HordeQBF unfairly since the majority of the benchmarks
is easy (solvable under a minute by DepQBF) and it makes no sense to use large computer 
clusters to solve them. In parallel computing, it is usual to
analyze the performance on many processors using \emph{weak scaling} where one
increases the amount of work involved in the considered instances proportionally
to the number of processors. Therefore in columns 7--9 we
restrict ourselves to ``big instances'' -- where DepQBF needs at least 10$\times$(the
number of cores used by HordeQBF) seconds 
to solve them. The average, total and median speedup values get significantly
larger and in fact we obtain highly superlinear average and median speedups.
Figure~\ref{fig_speedup} shows the distribution of speedups for these instances, it also reveals
how many instances (x-axis) qualify as ``big instances''.

\section{Conclusion}
We showed that QBF solving can be successfully parallelized using the same
techniques as for massively parallel SAT solving. Our parallel QBF solver
HordeQBF achieved superlinear total and median speedups for hard instances,
i.e., instances where parallelization makes sense.

As future work it would be interesting to consider further variants of
Q-resolution systems~\cite{DBLP:conf/sat/BalabanovWJ14} (apart from
traditional~\cite{DBLP:journals/iandc/BuningKF95} and long-distance
resolution~\cite{DBLP:journals/fmsd/BalabanovJ12,DBLP:conf/iccad/ZhangM02}) as
a means of diversification in HordeQBF, which would amount to a combination of
QBF proof systems with different power. Further, it may be promising to equip
HordeQBF with search space partitioning as
in \mbox{MPIDepQBF~\cite{DBLP:conf/sat/JordanKLS14}.}


\end{document}